%% file: main.tex
\documentclass[aps,prb,twocolumn,superscriptaddress,longbibliography]{revtex4-2}

\usepackage[colorlinks=true, linkcolor=blue, urlcolor=blue, citecolor=blue]{hyperref}
\usepackage{physics}
\usepackage{graphicx}
\usepackage{amsmath,amssymb}
\usepackage{bbm}
\usepackage{xcolor}
\usepackage{placeins}

\definecolor{revisionred}{rgb}{0.70,0.15,0.10}

\def\vk{\vb*{k}}
\def\vq{\vb*{q}}

\def\vR{\vb*{R}}

\begin{document}

\title{Single-orbital tight-binding model for chiral one-dimensional hybrid organic-inorganic lead halide perovskites}
\author{Yuya Ominato}
\email{ominato@aoni.waseda.jp}
\thanks{Contact author}
\affiliation{Waseda Institute for Advanced Study, Waseda University, Shinjuku-ku, Tokyo 169-0051, Japan}
\author{Tetsuya Furukawa}
\affiliation{Institute for Materials Research, Tohoku University, Sendai 980-8577, Japan}
\author{Ayumi Ishii}
\affiliation{Department of Chemistry and Biochemistry, School of Advanced Science and Engineering, Waseda University 3-4-1 Okubo, Shinjuku-ku, Tokyo 169-8555, Japan}
\author{Tetsuaki Itou}
\affiliation{Department of Applied Physics, Tokyo University of Science, Tokyo 125-8585, Japan}
\date{\today}

\begin{abstract}
We present a single-orbital tight-binding model for the low-energy electronic states of the chiral one-dimensional hybrid organic-inorganic lead halide perovskite \(\mathrm{(}R/S\mathrm{-PEA)PbI}_3\).
The model is constructed from a single effective orbital on each of the four symmetry-related sites in the primitive unit cell and incorporates layer, in-plane sublattice, and spin degrees of freedom.
With separate parameter sets for the conduction and valence bands, the effective Hamiltonian reproduces the overall band dispersions obtained from density-functional-theory calculations and quantitatively captures the spin splittings near the band edges.
It also captures the leading spin-polarization patterns of the Bloch states, showing that the band-edge spin splitting and spin polarization are encoded in a small number of symmetry-adapted spin-dependent hopping terms.
We further analyze the accidental degeneracies of the effective Hamiltonian using screw eigenvalues and antiunitary operators.
This analysis separates accidental degeneracies originating from the restricted term content of the effective Hamiltonian from degeneracies enforced by nonsymmorphic screw symmetries and time-reversal symmetry.
The present model provides a symmetry-transparent starting point for understanding the band-edge electronic structure of chiral lead halide perovskites and for analyzing optical, spin, and transport responses in this class of materials.
\end{abstract}

\maketitle

\section{Introduction}

Hybrid organic-inorganic perovskites provide a versatile platform for controlling low-dimensional electronic structures through the choice of organic cations and the resulting crystal symmetry \cite{Long2020,Pietropaolo2022}. In particular, chiral one-dimensional (1D) lead halide perovskites have attracted attention because they combine low-dimensional inorganic conducting frameworks, heavy elements with strong spin-orbit coupling, and intrinsically chiral crystal structures \cite{Ishii2025}.  The first chiral 1D organic-inorganic lead halide perovskite to be reported was \((S\)-PEA)\(\mathrm{PbBr}_3\), where PEA denotes 1-phenylethylamine, also referred to as methylbenzylamine (MBA) \cite{Billing2003}. Subsequent studies demonstrated chiroptical responses in related chiral lead iodide systems, including \((R/S\)-PEA)\(\mathrm{PbI}_3\), \((R/S\)-NEA)\(\mathrm{PbI}_3\), and \((R/S\)-CYHEA)\(\mathrm{PbI}_3\) \cite{Chen2019,Ishii2020,Hu2020}.  These developments established chiral lead halide perovskites as an important class of chiral semiconductors in which the inorganic framework and chiral organic molecules cooperatively determine the electronic and optical properties.

Recent studies have further revealed that the physical properties of chiral 1D lead halide perovskites are highly sensitive to crystal symmetry and chain arrangement. For example, \((R/S\)-NEA)\(\mathrm{PbI}_3\) has been reported to crystallize in both polar and nonpolar space groups, resulting in markedly different electronic and optical responses \cite{Ishii2025}. Density-functional-theory (DFT) band calculations have shown that the conduction-band minimum and valence-band maximum are mainly composed of Pb-\(6p\) and I-\(5p\) orbitals, respectively, and that the combination of strong spin-orbit coupling (SOC) and inversion-symmetry breaking gives rise to characteristic spin-split electronic structures near the band edges \cite{Wei2021,Xiao2024,Furukawa2026}. These findings indicate that the low-energy electronic structure near the band edges plays a central role in determining the optical, spin, and transport properties of chiral lead halide perovskites.

Despite these advances, a microscopic physical understanding of the low-energy electronic structure of chiral organic-inorganic lead halide perovskites remains limited. DFT calculations provide detailed information on the band structure, orbital character, and spin polarization, but they do not by themselves readily identify which low-energy degrees of freedom, symmetry-allowed hopping processes, and spin-dependent terms control these features. Therefore, a symmetry-adapted effective Hamiltonian is desirable for disentangling the relevant microscopic mechanisms and for developing theories of optical, spin, and transport responses. Such a model should retain the essential low-energy degrees of freedom near the band edges, respect the space-group symmetry, and reproduce the characteristic features of the DFT band structure. This is a nontrivial task because the unit cell contains a relatively complex organic-inorganic framework, and the effective hopping processes are determined not only by geometric bond lengths but also by orbital hybridization, the molecular environment, and downfolding effects from higher-energy states. To our knowledge, such a symmetry-adapted effective model has not yet been established for chiral 1D hybrid organic-inorganic lead halide perovskites.

In this work, we construct a single-orbital effective tight-binding model for a chiral 1D organic-inorganic lead halide perovskite using separate parameter sets for the conduction and valence bands.
The model is designed to describe the low-energy electronic states near the band edges, rather than to provide quantitative interpolation of the DFT band structure over a broad energy window.
Using the layer, in-plane sublattice, and spin degrees of freedom, we introduce spin-independent and spin-dependent hopping terms consistent with the space-group symmetry and time-reversal symmetry.
The resulting effective Hamiltonian reproduces the characteristic band structure obtained from DFT calculations and gives a quantitative description of the band dispersions near the band edges.
The model also captures the qualitative features of the spin polarization of the Bloch states.
These results indicate that the band-edge spin splitting and spin polarization are encoded in a simple symmetry-adapted tight-binding description.

The present model provides a transparent framework for understanding the electronic states of chiral organic-inorganic lead halide perovskites with complex crystal structures.
First-principles Wannier interpolation is generally more suitable for quantitatively describing the band structure and related matrix elements over a broad energy window.
In contrast, the present low-energy model identifies the essential symmetry-allowed hopping processes that govern the band-edge electronic structure.
In particular, the model enables us to relate the spin-dependent hopping terms to the band-edge spin splitting and spin-polarization patterns, providing microscopic insight into their origin.
Its compact form also reduces the computational cost of evaluating physical quantities, such as optical absorption and transport coefficients.
This effective-model approach complements DFT and Wannier-based methods as a symmetry-guided framework for understanding the band-edge electronic structure and for analyzing optoelectronic and spintronic functionalities in chiral lead halide perovskites.

The paper is organized as follows.
In Sec.~\ref{sec_model}, we introduce the symmetry-adapted effective tight-binding model for the chiral organic-inorganic lead halide perovskite with space group \(P2_12_12_1\) (No.~19).
We define the four-site basis, the layer and in-plane sublattice degrees of freedom, and the spin-independent and spin-dependent hopping terms of the effective Hamiltonian.
In Sec.~\ref{sec_band}, we compare the band structures obtained from the effective tight-binding model with those obtained from DFT calculations for the conduction and valence bands.
In Sec.~\ref{sec_spin}, we analyze the spin polarization of the corresponding Bloch states and compare the results with those from DFT calculations.
In Sec.~\ref{sec_lifting}, we separate accidental degeneracies of the effective Hamiltonian from symmetry-enforced degeneracies by adding additional spin-dependent terms.
In Sec.~\ref{sec_symmetry}, we analyze the remaining band degeneracies using screw and antiunitary operators along high-symmetry lines and at high-symmetry points.
Section~\ref{sec_conclusion} is devoted to the conclusion.

\begin{figure}
\begin{center}
\includegraphics[width=1\hsize]{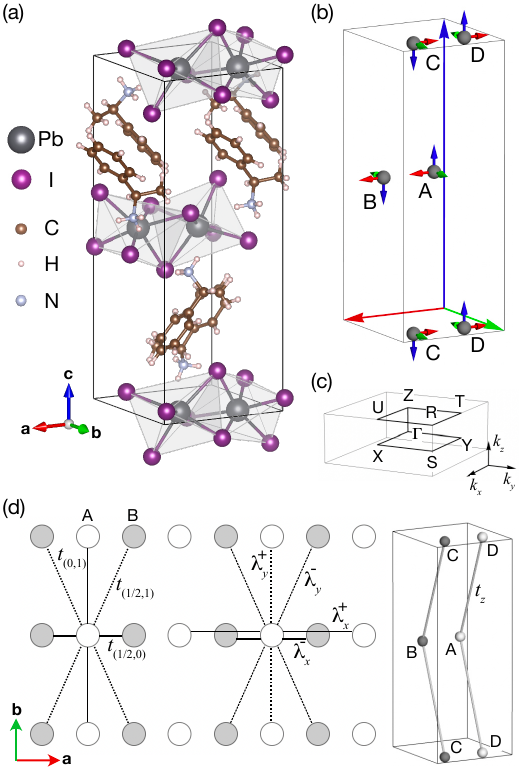}
\end{center}
\caption{
(a) Crystal structure of \(\mathrm{(}R\mathrm{-PEA)PbI}_3\).
(b) Minimal four-site lattice model with space group \(P2_12_12_1\), constructed from four symmetry-related effective sites occupying the \(4a\) Wyckoff position.
(c) Brillouin zone and high-symmetry lines.
(d) Schematic illustration of the hopping processes, including representative in-plane hopping within the AB plane and interlayer hopping between the AB and CD planes.
}
\label{fig_system}
\end{figure}

\section{Model}
\label{sec_model}

\begin{figure}
\begin{center}
\includegraphics[width=1\hsize]{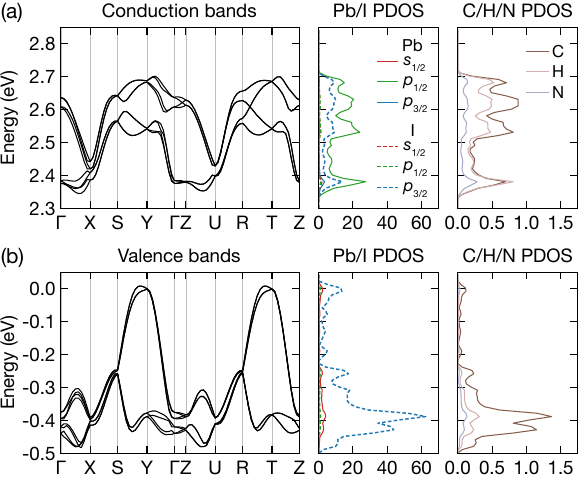}
\end{center}
\caption{
DFT energy bands and projected density of states (PDOS) for (a) the conduction bands and (b) the valence bands, plotted using the same DFT data as in Ref.~\cite{Furukawa2026}.
}
\label{fig_band_QE}
\end{figure}

We focus on \(\mathrm{(}R\mathrm{-PEA)PbI}_3\) as a representative chiral hybrid 1D organic-inorganic lead halide perovskite \cite{Chen2019}.
In this class of materials, the electronic structure can be tuned by selecting the chiral organic molecule, while the inorganic lead-iodide framework is largely retained.
As shown in Fig.~\ref{fig_system} (a), the crystal structure consists of chiral 1D lead-iodide chains running along the \(a\) axis, arranged into inorganic layers separated by chiral organic cations.
In this work, we use the DFT results reported in Ref.~\cite{Furukawa2026} as a reference for the low-energy electronic structure.
DFT calculations show that this material is an insulator with an energy gap of approximately \(2.3\,\mathrm{eV}\) \cite{Furukawa2026}, while the experimental energy gap has been estimated to be \(2.63\,\mathrm{eV}\) \cite{Makhija2025}.
The low-energy conduction and valence bands considered here are dominated by Pb and I orbital character, with a small contribution from the chiral organic molecules, as shown in Fig.~\ref{fig_band_QE} \cite{Furukawa2026}.
The DFT energy bands were obtained using the \textit{Quantum ESPRESSO} (QE) package~\cite{Giannozzi2009,Giannozzi2017}.
The generalized-gradient approximation was employed for the exchange-correlation functional \cite{Perdew1996}, together with a uniform \(8\times8\times8\) \(k\)-point mesh for the self-consistent-field loops. Further details of the DFT calculations can be found in Ref.~\cite{Furukawa2026}.

Informed by these DFT results, we construct an effective single-orbital model for the low-energy conduction and valence bands.
The effective orbital represents a low-energy degree of freedom mainly associated with the inorganic lead-iodide framework, while the effects of the surrounding organic molecules and higher-energy electronic states are absorbed into the hopping parameters.
The purpose of the effective model is to reproduce the characteristic low-energy band structure, spin splitting, and spin polarization obtained from the DFT calculations and to identify how these features arise from symmetry-allowed hopping processes.

In this section, we first introduce a minimal four-site lattice model for the low-energy degrees of freedom.
The model consists of four symmetry-related effective sites, representing the local chiral building blocks of the hybrid perovskite structure, occupying the \(4a\) Wyckoff position of \(P2_12_12_1\) (No.~19), as illustrated in Fig.~\ref{fig_system}(b).
We then specify the symmetry operations, introduce localized basis states on the four sites, and construct the corresponding symmetry-adapted tight-binding model.
After defining the Hamiltonian, we analyze the role of the spin-dependent hopping terms in the spin splitting.
The Brillouin zone and schematic illustration of hopping processes are shown in Figs.~\ref{fig_system}(c) and \ref{fig_system}(d), respectively, and the lattice constants used in this work are summarized in Table~\ref{table_lattice}.

\subsection{Lattice structure}

We consider a chiral 1D lead halide perovskite with the orthorhombic space group \(P2_12_12_1\).
The primitive lattice vectors are chosen as
\begin{align}
    \vb*{a}=a\vb*{e}_x,\quad
    \vb*{b}=b\vb*{e}_y,\quad
    \vb*{c}=c\vb*{e}_z,
\end{align}
and the corresponding reciprocal lattice vectors are
\begin{align}
    \vb*{a}^{\ast}=\frac{2\pi}{a}\vb*{e}_x,\quad
    \vb*{b}^{\ast}=\frac{2\pi}{b}\vb*{e}_y,\quad
    \vb*{c}^{\ast}=\frac{2\pi}{c}\vb*{e}_z,
\end{align}
where \(\vb*{e}_x\), \(\vb*{e}_y\), and \(\vb*{e}_z\) are orthonormal unit vectors.
The space group \(P2_12_12_1\) is generated by three twofold screw operations. 
In Seitz notation, these operations are written as
\begin{align}
    S_x
    &= \qty{C_{2x}\Big|\frac12\,\frac12\,0}, \notag \\
    S_y
    &= \qty{C_{2y}\Big|0\,\frac12\,\frac12}, \notag \\
    S_z
    &= \qty{C_{2z}\Big|\frac12\,0\,\frac12},
\end{align}
where the translational parts are expressed in fractional coordinates.
As the relevant sites of the effective tight-binding model, we introduce four symmetry-related effective sites in the primitive unit cell, denoted by \(A\), \(B\), \(C\), and \(D\).
These four sites constitute a particular realization of the \(4a\) Wyckoff position of the space group \(P2_12_12_1\).
Their fractional coordinates are
\begin{align}
    &(x_A,y_A,z_A) = \qty(\frac14,\frac14,\frac12),
    &(x_B,y_B,z_B) = \qty(\frac34,\frac14,\frac12), \notag \\
    &(x_C,y_C,z_C) = \qty(\frac34,\frac34,0),
    &(x_D,y_D,z_D) = \qty(\frac14,\frac34,0).
\end{align}
The corresponding real-space positions are
\begin{align}
    \vb*{\tau}_X
    =
    x_X\vb*{a}
    +
    y_X\vb*{b}
    +
    z_X\vb*{c},
\end{align}
where \(X\in \{A,B,C,D\}\).

\begin{table}[t]
    \centering
    \begin{tabular*}{\columnwidth}{@{\extracolsep{\fill}}lccc@{}}
        \hline\hline
        Compound & \(a\) (\AA) & \(b\) (\AA) & \(c\) (\AA) \\
        \hline
        \(\mathrm{(}R\mathrm{-PEA)PbI}_3\) & 8.0740 & 8.6460 & 20.862 \\
        \hline\hline
    \end{tabular*}
    \caption{Lattice constants.}
    \label{table_lattice}
\end{table}

\subsection{Symmetry-adapted tight-binding model}

We construct a symmetry-adapted tight-binding model for the low-energy electronic states near the band edges.
We introduce a single effective orbital on each of the four sites \(A\), \(B\), \(C\), and \(D\) in the unit cell.
The resulting four-dimensional site space is represented as a tensor product of layer and in-plane sublattice spaces.
We denote the Pauli matrices acting in these two spaces by \(\rho_i\) and \(\sigma_i\), respectively, and those acting in the spin space by \(s_i\), with \(i=1,2,3\), while \(i=0\) denotes the identity matrix in each space.
The Bloch Hamiltonian is written in the tensor-product space of the layer, sublattice, and spin spaces.
Throughout this paper, numerical subscripts are used for Pauli matrices, whereas Cartesian subscripts, such as \(x\), \(y\), and \(z\), are used for spatial directions in wave-vector components and hopping parameters of spin-dependent terms.
Products of matrices acting in different internal spaces are understood as Kronecker products.
For example, \(\rho_i\sigma_j s_l\) denotes \(\rho_i\otimes\sigma_j\otimes s_l\), where \(\rho_i\), \(\sigma_j\), and \(s_l\) act in the layer, sublattice, and spin spaces, respectively.
For clarity, identity matrices such as \(\rho_0\), \(\sigma_0\), and \(s_0\) are kept explicit in most expressions.

The Bloch basis is given by
\begin{align}
    \ket{\vk,X,s}
    =
    \frac{1}{\sqrt{N}}
    \sum_{\vR}
    e^{i\vk\cdot(\vR+\vb*{\tau}_{X})}
    \ket{\vR+\vb*{\tau}_{X},s},
    \label{eq_Bloch_basis}
\end{align}
where \(\vk\) is a wave vector in the Brillouin zone,  \(X\in \{A,B,C,D\}\) labels the four sites in the unit cell, \(s\in\{\uparrow,\downarrow\}\) labels the spin state in the \(s_3\) basis, and \(\vb*{\tau}_{X}\) is the position of site \(X\) in the unit cell.
Here, \(N\) is the number of unit cells, and \(\vR=m\vb*{a}+n\vb*{b}+p\vb*{c}\) is a Bravais lattice vector with \(m,n,p\in\mathbb{Z}\).
This convention, in which the site position is included in the Bloch phase, corresponds to the atomic gauge~\cite{Vanderbilt2018,EsteveParedes2023,WannierToolsAtomicGauge}.
In this gauge, a shift of the wave vector by a reciprocal lattice vector produces site-dependent phase factors, which are taken into account in the symmetry analysis in Sec.~\ref{sec_symmetry}.

With this basis, the Bloch Hamiltonian is represented as an \(8\times8\) matrix acting on the layer, in-plane sublattice, and spin spaces.
The effective tight-binding Hamiltonian consists of three terms,
\begin{align}
    H(\vk)
    =
    H_0(\vk)
    +V_x(\vk)
    +V_y(\vk),
\end{align}
where \(H_0(\vk)\) describes spin-independent hopping processes, while \(V_x(\vk)\) and \(V_y(\vk)\) describe spin-dependent hopping processes.
These spin-dependent terms are associated with SOC.
The first term \(H_0(\vk)\) preserves inversion symmetry, while \(V_x(\vk)\) and \(V_y(\vk)\) break inversion symmetry and reflect the chirality of the crystal. The distinction between right- and left-handed crystals is encoded in the signs of the parameters in \(V_x(\vk)\) and \(V_y(\vk)\).

The spin-independent part is given by
\begin{align}
    H_0(\vk)
    =
    h_0(\vk)\rho_0\sigma_0s_0
    +
    h_1(\vk)\rho_0\sigma_1s_0
    +
    h_z(\vk)\rho_1\sigma_1s_0,
    \label{eq_H0}
\end{align}
where the matrix elements are given by
\begin{align}
    h_0(\vk)
        =&\,
        \varepsilon_0
        +2t_{(1,0)}\cos\tilde{k}_x
        +2t_{(0,1)}\cos\tilde{k}_y
        \notag \\
        &
        +4t_{(1,1)}\cos\tilde{k}_x\cos\tilde{k}_y
        +2t_{(0,2)}\cos 2\tilde{k}_y, \label{eq_h0} \\
    h_1(\vk)
        =&\,
        2t_{(\frac12,0)}\cos\frac{\tilde{k}_x}{2}
        +2t_{(\frac32,0)}\cos\frac{3\tilde{k}_x}{2} \notag \\
        &
        +4t_{(\frac12,1)}
        \cos\frac{\tilde{k}_x}{2}
        \cos\tilde{k}_y
        +4t_{(\frac12,2)}
        \cos\frac{\tilde{k}_x}{2}
        \cos 2\tilde{k}_y, \label{eq_h1} \\
    h_z(\vk)
    =&\,
    4t_z
    \cos\frac{\tilde{k}_y}{2}
    \cos\frac{\tilde{k}_z}{2}.
    \label{eq_hz}
\end{align}
Here, we use the dimensionless wave numbers
\(\tilde{k}_x=k_xa\), \(\tilde{k}_y=k_yb\), and \(\tilde{k}_z=k_zc\).
The \(h_0(\vk)\) and \(h_1(\vk)\) terms originate from hopping processes within the same in-plane sublattice and between different in-plane sublattices, respectively.
The subscript of each hopping parameter denotes the corresponding in-plane displacement in units of \(a\) and \(b\).
The \(h_z(\vk)\) term couples the two layers and originates from interlayer hopping processes.

The spin-dependent part consists of two terms, \(V_x(\vk)\) and \(V_y(\vk)\), which involve \(s_1\) and \(s_2\), respectively.
The first term is given by
\begin{align}
    V_x(\vk)
    =
    2\lambda_x^{+}
    \sin\tilde{k}_x
    \rho_0\sigma_0s_1
    +
    2\lambda_x^{-}
    \sin\frac{\tilde{k}_x}{2}
    \rho_0\sigma_1s_1.
    \label{eq_Vx}
\end{align}
The two terms in \(V_x(\vk)\) describe spin-dependent hopping processes along the chiral chains.
The second term is given by
\begin{align}
    V_y(\vk)
    =
    2\lambda_y^{+}
    \sin\tilde{k}_y
    \rho_0\sigma_0s_2
    +
    4\lambda_y^{-}
    \cos\frac{\tilde{k}_x}{2}
    \sin\tilde{k}_y
    \rho_0\sigma_1s_2 .
    \label{eq_Vy}
\end{align}
These terms describe spin-dependent hopping processes between the chiral chains.
The corresponding hopping processes for \(\lambda_x^\pm\) and \(\lambda_y^\pm\) are schematically shown in Fig.~\ref{fig_system}(d).
The superscripts \(+\) and \(-\) distinguish terms that act with the same and opposite signs, respectively, in the \(\sigma_1=\pm1\) sublattice sectors.
The \(\lambda_x^{+}\) and \(\lambda_y^{+}\) terms have the \(\sigma_0\) matrix structure in the in-plane sublattice space and act with the same sign on the sublattice-even and sublattice-odd sectors.
By contrast, the \(\lambda_x^{-}\) and \(\lambda_y^{-}\) terms have the \(\sigma_1\) matrix structure and act with opposite signs on these two sectors.
This sublattice-sector dependence is essential for interpreting the spin splitting and spin polarization near the band edges.

The symmetry constraints on the Hamiltonian are expressed in terms of the matrix representations of the screw and time-reversal operations in the Bloch basis introduced above.
In the atomic gauge, the fractional translations associated with the screw operations enter their matrix representations as \(\vk\)-dependent phase factors.
These phase factors are included in the matrix representations of the screw operations, which are written as
\begin{align}
    \mathcal{S}_x(\vk)
    &=
    e^{-\mathrm{i}(\tilde{k}_x-\tilde{k}_y)/2}
    \rho_0\sigma_1(-\mathrm{i}s_1),
    \label{eq_screw_x_full}
    \\
    \mathcal{S}_y(\vk)
    &=
    e^{-\mathrm{i}(\tilde{k}_y-\tilde{k}_z)/2}
    \rho_1\sigma_0(-\mathrm{i}s_2),
    \label{eq_screw_y_full}
    \\
    \mathcal{S}_z(\vk)
    &=
    e^{-\mathrm{i}(\tilde{k}_z-\tilde{k}_x)/2}
    \rho_1\sigma_1(-\mathrm{i}s_3).
    \label{eq_screw_z_full}
\end{align}
The Hamiltonian satisfies
\begin{align}
    \mathcal{S}_i(\vk)H(\vk)\qty[\mathcal{S}_i(\vk)]^{-1}
    =
    H(C_{2i}\vk),
    \label{eq_screw_constraint_full}
\end{align}
where \(i=x,y,z\), and \(C_{2i}\) denotes the twofold rotation about the \(i\) axis.
The time-reversal operation is represented by
\begin{align}
    \Theta
    =
    \rho_0\sigma_0(-\mathrm{i}s_2)\mathcal{K},
    \label{eq_time_reversal}
\end{align}
where \(\mathcal{K}\) denotes complex conjugation.
This operator satisfies \(\Theta^2=-1\), and the Hamiltonian obeys
\begin{align}
    \Theta H(\vk)\Theta^{-1}
    =
    H(-\vk).
    \label{eq_time_reversal_constraint}
\end{align}
The effective Hamiltonian \(H(\vk)=H_0(\vk)+V_x(\vk)+V_y(\vk)\) satisfies these relations term by term and therefore respects both the \(P2_12_12_1\) space-group symmetry and time-reversal symmetry.

\subsection{Spin splitting along symmetry lines}

We first analyze the spin-independent Hamiltonian \(H_0(\vk)\) in Eq.~\eqref{eq_H0} before discussing the spin splitting induced by \(V_x(\vk)\)  and \(V_y(\vk)\).
The matrices \(\rho_0\sigma_0s_0\), \(\rho_0\sigma_1s_0\), and \(\rho_1\sigma_1s_0\) commute with one another.
Therefore, the eigenstates of \(H_0(\vk)\) can be labeled by the eigenvalues
\(\mu=\pm1\) and \(\nu=\pm1\) of \(\rho_1\) and \(\sigma_1\), respectively.
Here, \(\mu\) labels the layer-even/odd sector and \(\nu\) labels the sublattice-even/odd sector.
The corresponding eigenvalues of the spin-independent part are
\begin{align}
    E_{\mu\nu}^{(0)}(\vk)
    =
    h_0(\vk)
    +
    \nu h_1(\vk)
    +
    \mu\nu h_z(\vk),
\end{align}
with a twofold spin degeneracy.
Thus, \(h_1(\vk)\) separates the sublattice-even and sublattice-odd sectors, while \(h_z(\vk)\) further splits the layer-even and layer-odd sectors.
Since \(h_z(\vk)\) in Eq.~\eqref{eq_hz} vanishes on the Brillouin-zone boundary planes \(\tilde{k}_y=\pi\) and \(\tilde{k}_z=\pi\), the spin-independent model has an additional degeneracy between the layer-even and layer-odd sectors on these planes.

The spin-dependent terms in \(V_x(\vk)\) and \(V_y(\vk)\) lead to different spin-splitting patterns in the two sublattice sectors.
The \(\lambda_x^{+}\) and \(\lambda_y^{+}\) terms are proportional to \(\rho_0\sigma_0s_1\) and \(\rho_0\sigma_0s_2\), respectively.
Because their sublattice part is \(\sigma_0\), these terms give the same spin splitting in the \(\nu=+1\) and \(\nu=-1\) sectors.
By contrast, the \(\lambda_x^{-}\) and \(\lambda_y^{-}\) terms are proportional to \(\rho_0\sigma_1s_1\) and \(\rho_0\sigma_1s_2\), respectively.
Because their sublattice part is \(\sigma_1\), the effective spin-dependent field changes sign between the two sectors.
As a result, the \(\lambda_x^{+}\) and \(\lambda_y^{+}\) terms preserve the relative ordering of the two spin branches between the sublattice-even and sublattice-odd sectors, while the \(\lambda_x^{-}\) and \(\lambda_y^{-}\) terms reverse this ordering.
This distinction gives rise to qualitatively different spin-splitting patterns produced by the sublattice-even and sublattice-odd spin-dependent hopping terms.

This sublattice-sector dependence is reflected in the spin splitting along the symmetry lines.
Along the \(\Gamma\)-X line, the relevant spin-dependent terms are the \(s_1\)-dependent terms in \(V_x(\vk)\).
The \(\lambda_x^{+}\) term is proportional to \(\rho_0\sigma_0s_1\) and produces the same \(s_1\) splitting in the \(\nu=+1\) and \(\nu=-1\) sectors.
By contrast, the \(\lambda_x^{-}\) term is proportional to \(\rho_0\sigma_1s_1\) and produces opposite \(s_1\) splittings in the two sectors.
Thus, the relative magnitude and sign of \(\lambda_x^{+}\) and \(\lambda_x^{-}\) determine the ordering of the spin-resolved energy bands along \(\Gamma\)-X.
The same reasoning applies to the \(\Gamma\)-Y line, where the relevant spin-dependent terms are the \(s_2\)-dependent terms in \(V_y(\vk)\).
The \(\lambda_y^{+}\) term is proportional to \(\rho_0\sigma_0s_2\) and produces the same \(s_2\) splitting in the \(\nu=+1\) and \(\nu=-1\) sectors.
By contrast, the \(\lambda_y^{-}\) term is proportional to \(\rho_0\sigma_1s_2\) and produces opposite \(s_2\) splittings in the two sectors.
Thus, the spin-polarization patterns along \(\Gamma\)-X and \(\Gamma\)-Y are understood from the same sublattice-sector classification, with \(s_1\) and \(s_2\) as the relevant spin components, respectively.

The effective Hamiltonian captures the overall band dispersion and the spin splitting.
The weak \(k_z\) dispersion is controlled by the interlayer hopping term \(h_z(\vk)\), and the spin splitting is governed by the spin-dependent hopping parameters.
Because the effective Hamiltonian retains only a restricted set of symmetry-adapted terms designed to reproduce the overall band dispersion and the spin splitting near the band edges, some accidental degeneracies remain along high-symmetry lines and at high-symmetry points.
These degeneracies are not enforced by the \(P2_12_12_1\) space-group symmetry.
Instead, they originate from the restricted term content of the effective Hamiltonian.
In Sec.~\ref{sec_lifting}, we show that these accidental degeneracies can be lifted by additional symmetry-adapted spin-dependent terms.

\section{Energy band structure}
\label{sec_band}

\begin{table}
    \centering
    \begin{tabular*}{\columnwidth}{@{\extracolsep{\fill}}cccc@{}}
        \hline\hline
        Parameter & Conduction bands & Valence bands \\
        \hline
        \(\varepsilon_0\) & 2.551 & -0.312 \\
        \(t_{(\frac12,0)}\) & 0.041 & 0.075 \\
        \(t_{(1,0)}\) & 0.012 & 0.008 \\
        \(t_{(\frac32,0)}\) & -0.004 & -0.021 \\
        \(t_{(0,1)}\) & -0.038 & -0.044 \\
        \(t_{(\frac12,1)}\) & 0.006 & -0.021 \\
        \(t_{(1,1)}\) & 0.005 & -0.004 \\
        \(t_{(0,2)}\) & -0.008 & 0.000 \\
        \(t_{(\frac12,2)}\) & 0.005 & 0.000 \\
        \(t_z\) & 0.002 & 0.002 \\
        \(\lambda_x^{+}\) &  0.002 & -0.005  \\
        \(\lambda_x^{-}\) & -0.005 & -0.001  \\
        \(\lambda_y^{+}\) & -0.001 &  0.001 \\
        \(\lambda_y^{-}\) & -0.004 & -0.002  \\
        \hline\hline
    \end{tabular*}
    \caption{
    Hopping parameters for the conduction- and valence-band sectors.
    All parameters are given in units of eV.
    }
    \label{table_tb_parameters}
\end{table}

\begin{figure*}
\begin{center}
\includegraphics[width=1\hsize]{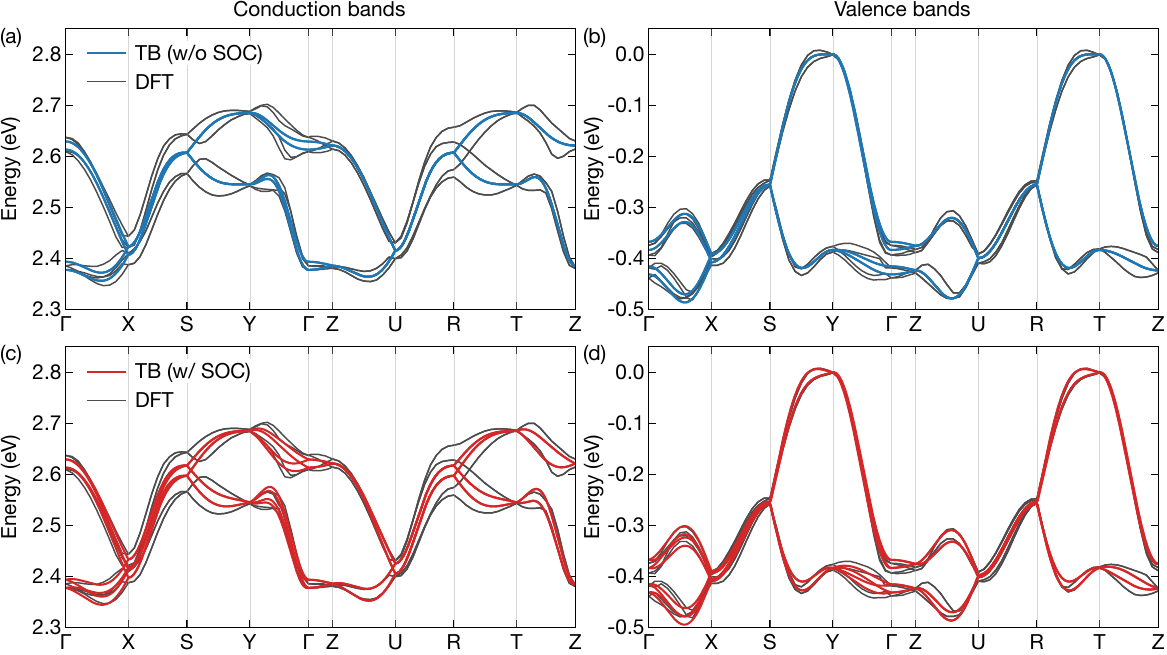}
\end{center}
\caption{
Comparison of the band structures obtained from the symmetry-adapted tight-binding model and from DFT calculations using QE.
Colored and gray curves denote the tight-binding (TB) and DFT results, respectively.
Panels (a) and (b) show the conduction and valence bands obtained from \(H_0(\vk)\). Panels (c) and (d) show the corresponding bands obtained from \(H_0(\vk)+V_x(\vk)+V_y(\vk)\).
}
\label{fig_band}
\end{figure*}

Figure~\ref{fig_band} compares the tight-binding band structures with the DFT band structures for the conduction- and valence-band sectors.
The energy is measured relative to the highest valence-band energy at the Y point.
We first examine the role of the spin-independent part \(H_0(\vk)\).
With separate parameter sets for the conduction and valence bands, as listed in Table~\ref{table_tb_parameters}, \(H_0(\vk)\) reproduces the overall dispersion of the corresponding DFT bands, as shown in Figs.~\ref{fig_band}(a) and \ref{fig_band}(b).
This agreement shows that the spin-independent hopping processes in \(H_0(\vk)\) capture the overall shape of the low-energy band dispersion within the effective tight-binding description.

The spin-dependent terms \(V_x(\vk)\) and \(V_y(\vk)\) are included to reproduce the spin splittings near the band edges.
These terms split the energy bands without substantially changing the overall dispersion obtained from \(H_0(\vk)\).
The tight-binding results quantitatively agree with the DFT results near the band edges, reproducing both the energy dispersions and the spin splittings, as shown in Figs.~\ref{fig_band}(c) and \ref{fig_band}(d).
Thus, the effective Hamiltonian provides a reasonable low-energy description of the conduction- and valence-band edges.

The fitted hopping parameters listed in Table~\ref{table_tb_parameters} show distinct trends between the conduction and valence bands.
For \(V_x(\vk)\), the \(\lambda_x^{-}\) term gives the main contribution in the conduction bands, whereas the \(\lambda_x^{+}\) term gives the main contribution in the valence bands.
The conduction-band edge lies on the \(\Gamma\)-X line, where the spin splitting is well described by the nearest-neighbor spin-dependent hopping term along the chiral chains, \(2\lambda_x^{-}\sin\frac{\tilde{k}_x}{2}\rho_0\sigma_1s_1\).
By contrast, the valence-band edge lies on the S-Y line, where the spin splitting is sizable away from the endpoints and strongly suppressed near the S and Y points.
This behavior is captured by the spin-dependent hopping term \(2\lambda_x^{+}\sin\tilde{k}_x\rho_0\sigma_0s_1\), which vanishes at both endpoints of the S-Y line.
For \(V_y(\vk)\), the \(\lambda_y^{-}\) term gives the main contribution in both the conduction and valence bands.
The role of this term will be discussed in connection with the spin-polarization pattern in Sec.~\ref{sec_spin}.

Away from the band extrema, the tight-binding bands also reproduce the overall trend of the DFT bands.
Thus, the effective Hamiltonian captures the target features of the DFT results, namely the overall band dispersion and the spin splittings near the band edges.
At the same time, some quantitative deviations remain away from the band edges, as expected for a simple low-energy model with a restricted set of symmetry-adapted terms.
Furthermore, the tight-binding bands retain additional degeneracies along the \(\Gamma\)-Z and U-R lines, whereas the corresponding DFT bands are split along these lines.
These residual degeneracies are not required by the \(P2_12_12_1\) space-group symmetry and reflect the restricted term content of the effective Hamiltonian.

\section{Spin polarization}
\label{sec_spin}

\begin{figure}
\begin{center}
\includegraphics[width=1\hsize]{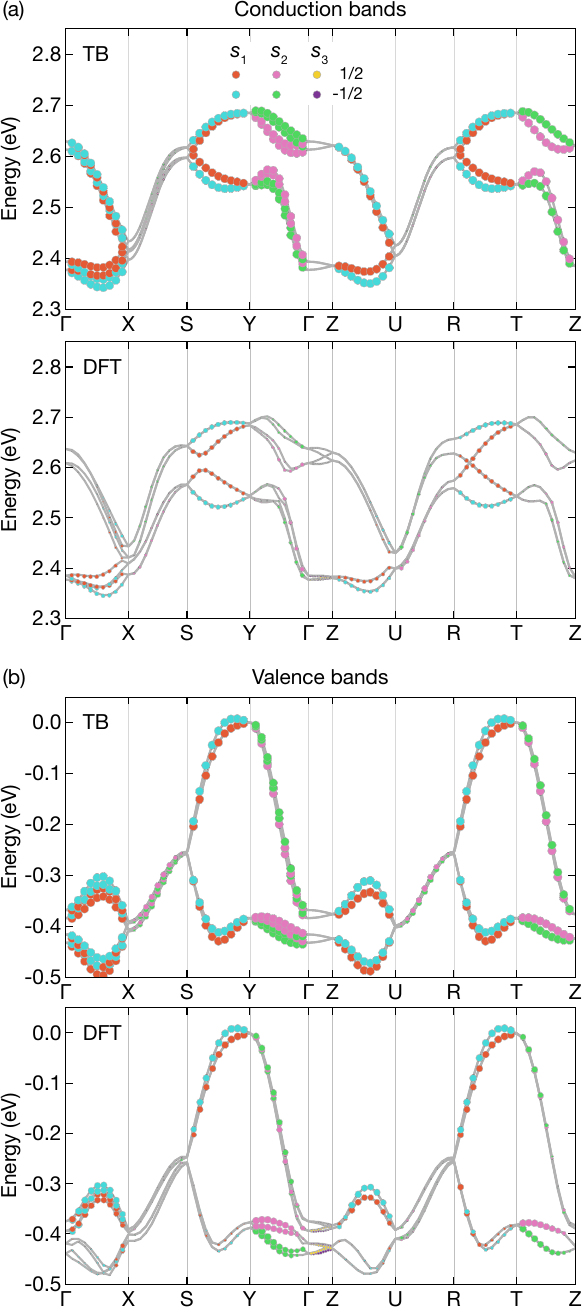}
\end{center}
\caption{
Spin polarization of the conduction and valence bands obtained from the effective tight-binding Hamiltonian and from DFT.
The spin expectation values are plotted on the corresponding energy bands.
(a) Conduction bands.
(b) Valence bands.
For each panel, the upper and lower plots show the tight-binding (TB) and DFT results, respectively.
}
\label{fig_spin}
\end{figure}

We next compare the spin polarization obtained from the effective tight-binding Hamiltonian with that obtained from the DFT calculations.
In the presence of degeneracies, the spin polarization is evaluated by averaging the spin operator over the degenerate subspace.
Along each high-symmetry line, we show only the spin component that can remain finite after this averaging procedure.
Specifically, we plot \(s_1\) on \(\Gamma\)-X, S-Y, Z-U, and R-T, \(s_2\) on X-S, Y-\(\Gamma\), U-R, and T-Z, and \(s_3\) on \(\Gamma\)-Z.
The symmetry constraints determining which spin components can remain finite after the averaging procedure are discussed in Sec.~\ref{sec_symmetry} using the corresponding screw operators.
Figure~\ref{fig_spin} shows the resulting spin polarization on the energy bands for the conduction and valence bands.
For the tight-binding model, the spin polarization is evaluated from the expectation values of \(\rho_0\sigma_0s_i/2\), and the DFT spin polarization is obtained from the spin expectation values of the Kohn-Sham Bloch states.
In the plots, the colored markers are overlaid on the energy bands, with the marker color representing the sign of the spin expectation value and the marker size representing its magnitude.

The tight-binding model qualitatively reproduces the spin-polarization patterns obtained from DFT in both the conduction and valence bands.
Near the band edges, their sign patterns are captured by the tight-binding model.
Away from the band edges, the DFT results show a tendency toward suppressed spin polarization, and quantitative deviations from the tight-binding results become more visible.
Thus, the model should be regarded as an effective description of the leading spin-polarization structure near the band edges rather than as a fully quantitative description of the spin polarization over the energy range considered here.

For the conduction bands, the tight-binding model reproduces the sign pattern of the DFT spin polarization for both the \(s_1\) and \(s_2\) components, as shown in Fig.~\ref{fig_spin}(a).
Along the lines where \(s_1\) is plotted, the spin polarization follows the pattern expected from the \(\lambda_x^{-}\) term in \(V_x(\vk)\). The spin polarization changes sign between the two sublattice sectors.
The same type of sign pattern is seen for the \(s_2\) component on the corresponding lines, reflecting the role of the \(\lambda_y^{-}\) term in \(V_y(\vk)\).
On the \(\Gamma\)-Z line, the \(s_3\) component vanishes in the tight-binding model because the effective Hamiltonian contains no \(s_3\)-dependent term, and the DFT results show negligible \(s_3\) polarization.

For the valence bands, the tight-binding model captures the main qualitative features of the DFT spin-polarization pattern, as shown in Fig.~\ref{fig_spin}(b).
The \(s_2\) component shows a consistent sign pattern between the tight-binding and DFT results on the corresponding high-symmetry lines.
This behavior reflects the role of the \(\lambda_y^{-}\) term in \(V_y(\vk)\), which is needed to reproduce the ordering of the spin-polarized branches in the valence bands.
For the \(s_1\) component, the band-edge spin polarization is mainly described by the \(\lambda_x^{+}\) term in \(V_x(\vk)\), consistent with the spin splitting along the S-Y line discussed above.
Away from the valence-band edge, however, the DFT spin polarization tends to be suppressed, and the tight-binding model does not reproduce its quantitative behavior.

The spin-polarization analysis shows that the effective Hamiltonian captures the main spin components and branch-dependent sign patterns near the conduction- and valence-band edges.
Within the effective model, simultaneous sign reversal of \(\lambda^\pm_x\) and \(\lambda^\pm_y\) reverses the spin-polarization pattern, corresponding to the opposite crystal chirality \cite{Furukawa2026}.
Quantitative differences remain, especially away from the band edges, where the tight-binding model tends to overestimate some spin-polarization amplitudes and does not reproduce all details of the DFT spin polarization.
These deviations reflect the restricted set of spin-dependent hopping terms retained in the effective Hamiltonian.

\section{Lifting accidental degeneracies}
\label{sec_lifting}

\begin{figure}
\begin{center}
\includegraphics[width=1\hsize]{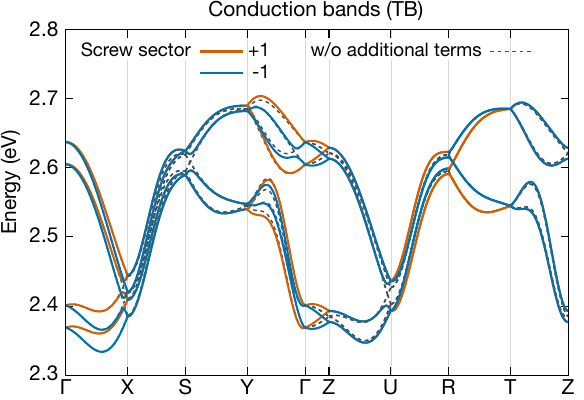}
\end{center}
\caption{
Band structure of the extended Hamiltonian.
Solid and dashed curves denote the results with and without the additional spin-dependent terms, respectively.
}
\label{fig_band_waSOC}
\end{figure}

\begin{table}
    \centering
    \begin{tabular*}{\columnwidth}{@{\extracolsep{\fill}}cccccc@{}}
        \hline\hline
        \(V_x(\vk),V_y(\vk)\) &  & \(V_1(\vk)\) &  & \(V_2(\vk)\) &  \\
        \hline
        \(\lambda_x^{+}\) & 0.002
        & \(\lambda_y^{\mathrm{Y}}\) & -0.001
        & \(\lambda_z^{\Gamma\mathrm{X}}\) & -0.001 \\
        \(\lambda_x^{-}\) & -0.006
        & \(\lambda_z^{\mathrm{Z}}\) & -0.002
        & \(\lambda_z^{\mathrm{Y}\Gamma}\) & -0.006 \\
        \(\lambda_y^{+}\) & -0.002
        & \(\lambda_z^{\mathrm{R}}\) & -0.001
        & \(\lambda_x^{\mathrm{SY}}\) & -0.003 \\
        \(\lambda_y^{-}\) & -0.006
        & 
        & 
        & \(\lambda_x^{\mathrm{ZU}}\) & -0.002 \\
        \hline\hline
    \end{tabular*}
    \caption{
    Parameters used to calculate the energy bands in Fig.\,\ref{fig_band_waSOC}. The other parameters are the same as those used for the conduction bands in Table~\ref{table_tb_parameters}, except that \(t_z=0.004\).
    }
    \label{table_extended_soc_parameters}
\end{table}

In this section, we show that the accidental degeneracies of the effective Hamiltonian can be removed by extending the Hamiltonian with additional symmetry-adapted spin-dependent terms.
As shown in the previous sections, the effective Hamiltonian captures the band-edge dispersion and the leading spin splittings, but still retains degeneracies that are not enforced by the \(P2_12_12_1\) space-group symmetry (No.~19).
These accidental degeneracies originate from the restricted set of symmetry-adapted terms retained in the effective Hamiltonian.
The purpose of the extension introduced below is not to further improve the quantitative fit to the DFT bands, but to show that these accidental degeneracies can be lifted while preserving the space-group symmetry and time-reversal symmetry.
By adding the additional terms, we obtain an extended Hamiltonian in which accidental degeneracies are lifted and the corresponding band crossings become anticrossings, leaving only the degeneracies enforced by nonsymmorphic screw symmetries and time-reversal symmetry.

The additional spin-dependent terms are given by
\begin{align}
    V_1(\vk)
    =&
    4\lambda_y^{\mathrm{Y}}
    \sin\frac{\tilde{k}_y}{2}
    \cos\frac{\tilde{k}_z}{2}
    \rho_1\sigma_1s_2 \notag \\
    &+
    4\lambda_z^{\mathrm{Z}}
    \cos\frac{\tilde{k}_y}{2}
    \sin\frac{\tilde{k}_z}{2}
    \rho_1\sigma_1s_3 \notag \\
    &+
    4\lambda_z^{\mathrm{R}}
    \sin\frac{\tilde{k}_y}{2}
    \sin\frac{\tilde{k}_z}{2}
    \rho_1\sigma_2s_3 ,
\end{align}
and
\begin{align}
    V_2(\vk)
    =&
    8\lambda_z^{\mathrm{\Gamma X}}
    \sin\frac{\tilde{k}_x}{2}
    \cos\frac{\tilde{k}_y}{2}
    \cos\frac{\tilde{k}_z}{2}
    \rho_1\sigma_3s_3 \notag \\
    &+
    4\lambda_x^{\mathrm{SY}}
    \sin\frac{\tilde{k}_y}{2}
    \cos\frac{\tilde{k}_z}{2}
    \rho_2\sigma_2s_1 \notag \\
    &+
    2\lambda_z^{\mathrm{Y\Gamma}}
    \sin\tilde{k}_y
    \rho_3\sigma_0s_3 \notag \\
    &+
    8\lambda_x^{\mathrm{ZU}}
    \sin\frac{\tilde{k}_x}{2}
    \cos\frac{\tilde{k}_y}{2}
    \sin\frac{\tilde{k}_z}{2}
    \rho_2\sigma_3s_1 .
\end{align}
These terms preserve both the \(P2_12_12_1\) space-group symmetry and time-reversal symmetry.
They are introduced not as a further quantitative fit to the DFT bands, but as a symmetry-adapted extension used to lift accidental degeneracies of the effective Hamiltonian.
In the extended Hamiltonian, accidental degeneracies can be removed by the additional spin-dependent terms.
Degeneracies enforced by nonsymmorphic screw symmetries and time-reversal symmetry remain protected.
This distinction allows us to identify which band crossings are symmetry protected and which become anticrossings under symmetry-adapted perturbations.

Figure~\ref{fig_band_waSOC} shows the band structure of the extended Hamiltonian,
\begin{align}
    H(\vk)
    =
    H_0(\vk)
    +V_x(\vk)
    +V_y(\vk)
    +V_1(\vk)
    +V_2(\vk).
\end{align}
The solid curves show the energy bands obtained from the extended Hamiltonian including \(V_1(\vk)\) and \(V_2(\vk)\). The dashed curves show the results obtained by setting \(V_1(\vk)=V_2(\vk)=0\), while keeping all other parameters identical to those used for the solid curves.
The hopping parameters for the spin-dependent terms are listed in Table~\ref{table_extended_soc_parameters}.
They are not set to improve the quantitative agreement with the DFT bands.
Instead, their values are chosen to make the lifting of accidental degeneracies and the formation of anticrossings clearly visible.
For this demonstration, we focus on the conduction bands, where the relevant degeneracy structure and the effects of the additional terms can be seen clearly.

The role of the additional terms is summarized in Tables~\ref{table_degeneracy} and \ref{table_high_symm_eigenvalues}.
In Table~\ref{table_degeneracy}, \(d\) denotes the degeneracy of the effective Hamiltonian, while \(g\) denotes that of the extended Hamiltonian.
The comparison between \(d\) and \(g\) shows that the effective Hamiltonian contains additional degeneracies that are not enforced by the \(P2_12_12_1\) space-group symmetry or time-reversal symmetry.
After the additional terms are included, the remaining degeneracies are consistent with those expected from the standard representation theory of space groups with time-reversal symmetry~\cite{Herring1937,Bradley2010} and with band sticking in nonsymmorphic crystals~\cite{Konig1997,Konig1999}.
Table~\ref{table_high_symm_eigenvalues} gives the eigenvalues of the extended Hamiltonian at the high-symmetry points and shows which additional parameters control the splittings at each point.

The terms in \(V_1(\vk)\) lift the accidental degeneracies at the high-symmetry points, while the terms in \(V_2(\vk)\) hybridize accidentally crossing branches on the high-symmetry lines and open gaps.
At the same time, the degeneracies enforced by nonsymmorphic screw symmetries and time-reversal symmetry remain intact.
Therefore, the extended Hamiltonian provides a convenient setting for separating accidental degeneracies of the effective Hamiltonian from symmetry-enforced degeneracies.

\begin{table}[t]
    \centering
    \begin{tabular*}{\columnwidth}{@{\extracolsep{\fill}}cccccc@{}}
        \hline
        \hline
        Symmetry point & \(d\) & \(g\) & Symmetry line & \(d\) & \(g\) \\
        \hline
        \(\Gamma\)
        & 2
        & 2
        & \(\Gamma\)--X
        & 1
        & 1
        \\
        \(\mathrm{X}\)
        & 2
        & 2
        & X--S
        & 2
        & 2
        \\
        \(\mathrm{S}\)
        & 4
        & 4
        & S--Y
        & 2
        & 2
        \\
        \(\mathrm{Y}\)
        & 4
        & 2
        & Y--\(\Gamma\)
        & 1
        & 1
        \\
        \(\mathrm{Z}\)
        & 4
        & 2
        & \(\Gamma\)--Z
        & 2
        & 1
        \\
        \(\mathrm{U}\)
        & 4
        & 4
        & Z--U
        & 2
        & 2
        \\
        \(\mathrm{R}\)
        & 4
        & 2
        & U--R
        & 4
        & 2
        \\
        \(\mathrm{T}\)
        & 4
        & 4
        & R--T
        & 2
        & 2
        \\
        &
        &
        &
        T--Z
        & 2
        & 2
        \\
        \hline
        \hline
    \end{tabular*}
    \caption{
    Degeneracies at the high-symmetry points and along the high-symmetry lines.
    \(d\) and \(g\) denote the degeneracy of the effective Hamiltonian and the extended Hamiltonian, respectively.
    }
    \label{table_degeneracy}
\end{table}

\begin{table}[t]
    \centering
    \begin{tabular*}{\columnwidth}{@{\extracolsep{\fill}}cc@{}}
        \hline
        \hline
        Symmetry point & Eigenvalue \\
        \hline
        \(\Gamma\)
        &
        \(\begin{aligned}
        E_{\eta,\zeta}^{\Gamma}
        &=
        \varepsilon^{\Gamma}
        +\eta \Delta^{\Gamma}
        +4\zeta t_z
        \end{aligned}\)
        \\
        \(\mathrm{X}\)
        &
        \(\begin{aligned}
        E_{\eta,\zeta}^{\rm X}
        &=
        \varepsilon^{\rm X}
        +2\eta\lambda_x^{-}
        +4\zeta
        \sqrt{
            t_z^2+\qty(2\lambda_z^{\Gamma{\rm X}})^2
        }
        \end{aligned}\)
        \\
        \(\mathrm{S}\)
        &
        \(\begin{aligned}
        E_{\eta}^{\rm S}
        &=
        \varepsilon^{\rm S}
        +2\eta
        \sqrt{
            \qty(\lambda_x^{-})^2
            +\qty(2\lambda_x^{\mathrm{SY}})^2
            +\qty(2\lambda_y^{\mathrm{Y}})^2
        }
        \end{aligned}\)
        \\
        \(\mathrm{Y}\)
        &
        \(\begin{aligned}
        E_{\eta,\zeta}^{\rm Y}
        &=
        \varepsilon^{\rm Y}
        +\eta
        \sqrt{
            \qty(\Delta^{\rm Y}+4\zeta\lambda_y^{\mathrm{Y}})^2
            +\qty(4\lambda_x^{\mathrm{SY}})^2
        }
        \end{aligned}\)
        \\
        \(\mathrm{Z}\)
        &
        \(\begin{aligned}
        E_{\eta,\zeta}^{\rm Z}
        &=
        \varepsilon^{\Gamma}
        +\eta \Delta^{\Gamma}
        +4\zeta\lambda_z^{\mathrm{Z}}
        \end{aligned}\)
        \\
        \(\mathrm{U}\)
        &
        \(\begin{aligned}
        E_{\eta}^{\rm U}
        &=
        \varepsilon^{\rm X}
        +2\eta
        \sqrt{
            \qty(\lambda_x^{-})^2
            +\qty(4\lambda_x^{\mathrm{ZU}})^2
            +\qty(2\lambda_z^{\mathrm{Z}})^2
        }
        \end{aligned}\)
        \\
        \(\mathrm{R}\)
        &
        \(\begin{aligned}
        E_{\eta,\zeta}^{\rm R}
        &=
        \varepsilon^{\rm S}
        +2\eta\lambda_x^{-}
        +4\zeta\lambda_z^{\mathrm{R}}
        \end{aligned}\)
        \\
        \(\mathrm{T}\)
        &
        \(\begin{aligned}
        E_{\eta}^{\rm T}
        &=
        \varepsilon^{\rm Y}
        +\eta
        \sqrt{
            \qty(\Delta^{\rm Y})^2
            +\qty(4\lambda_z^{\mathrm{R}})^2
        }
        \end{aligned}\)
        \\
        \hline
        \hline
    \end{tabular*}

    \vspace{1ex}

    \begin{equation}
    \begin{aligned}
    \varepsilon^{\Gamma}
    &=
    \varepsilon_0
    +2\qty[
        t_{(1,0)}
        +t_{(0,1)}
        +2t_{(1,1)}
        +t_{(0,2)}
    ] \\
    \varepsilon^{\rm X}
    &=
    \varepsilon_0
    -2\qty[
        t_{(1,0)}
        -t_{(0,1)}
        +2t_{(1,1)}
        -t_{(0,2)}
    ] \\
    \varepsilon^{\rm S}
    &=
    \varepsilon_0
    -2\qty[
        t_{(1,0)}
        +t_{(0,1)}
        -2t_{(1,1)}
        -t_{(0,2)}
    ] \\
    \varepsilon^{\rm Y}
    &=
    \varepsilon_0
    +2\qty[
        t_{(1,0)}
        -t_{(0,1)}
        -2t_{(1,1)}
        +t_{(0,2)}
    ] \\
    \Delta^{\Gamma}
    &=
    2\qty[
        t_{(\frac12,0)}
        +t_{(\frac32,0)}
        +2\qty(
            t_{(\frac12,1)}
            +t_{(\frac12,2)}
        )
    ] \\
    \Delta^{\rm Y}
    &=
    2\qty[
        t_{(\frac12,0)}
        +t_{(\frac32,0)}
        -2\qty(
            t_{(\frac12,1)}
            -t_{(\frac12,2)}
        )
    ] \notag
    \end{aligned}
    \end{equation}

    \caption{
    Eigenvalues of the extended Hamiltonian at the high-symmetry points, where \(\eta=\pm1\) and \(\zeta=\pm1\).
    }
    \label{table_high_symm_eigenvalues}
\end{table}

\section{Symmetry analysis}
\label{sec_symmetry}

We analyze the remaining band degeneracies using the matrix representations of the screw and time-reversal operations.
The analysis is based on the algebra of these symmetry operations and explains how the nonsymmorphic screw symmetries and time-reversal symmetry enforce the degeneracies summarized in Table~\ref{table_degeneracy}.
Related analyses based on screw eigenvalues for space group \(P2_12_12_1\) (No.~19) have been discussed in Refs.~\cite{Furusaki2017,Chang2018,Leonhardt2021}.
We first classify the Bloch states on high-symmetry lines by the eigenvalues of the corresponding screw operators.
We then combine the screw operations with the time reversal operation to identify the Kramers degeneracies on the high-symmetry lines and the fourfold degeneracies at the high-symmetry points.
The same symmetry framework is also used to determine which spin components can remain finite after averaging over a degenerate subspace.

We first construct the screw operators used to label the Bloch states on the high-symmetry lines.
In the Bloch basis introduced in Sec.~\ref{sec_model}, a shift of the wave vector by a reciprocal lattice vector \(\vb*{G}\) gives
\begin{align}
    \ket{\vk+\vb*{G},X,s}
    =
    e^{\mathrm{i}\vb*{G}\cdot\vb*{\tau}_{X}}
    \ket{\vk,X,s}.
\end{align}
We introduce the corresponding transformation
\begin{align}
    U(\vb*{G})
    =
    \mathrm{diag}
    \left(
        e^{\mathrm{i}\vb*{G}\cdot\vb*{\tau}_{A}},
        e^{\mathrm{i}\vb*{G}\cdot\vb*{\tau}_{B}},
        e^{\mathrm{i}\vb*{G}\cdot\vb*{\tau}_{C}},
        e^{\mathrm{i}\vb*{G}\cdot\vb*{\tau}_{D}}
    \right)
    \otimes s_0.
\end{align}
The Bloch Hamiltonian satisfies
\begin{align}
    H(\vk+\vb*{G})
    =
    [U(\vb*{G})]^{-1}
    H(\vk)
    U(\vb*{G}) .
    \label{eq_atomic_gauge_periodicity}
\end{align}
We consider a wave vector \(\vq\) on a high-symmetry line.
For the relevant screw operation on this line, the rotational part satisfies
\begin{align}
    C_{2i}\vq
    =
    \vq+\vb*{G}_{\mathrm{S}},
\end{align}
where \(i=x,y,z\), and \(\vb*{G}_{\mathrm{S}}\) is the reciprocal lattice vector associated with the high-symmetry line and screw operation considered.
Together with Eqs.~\eqref{eq_screw_constraint_full} and \eqref{eq_atomic_gauge_periodicity}, this relation leads to a unitary operator acting within the space of Bloch states at \(\vq\)
\begin{align}
    \mathcal{S}^{\mathrm{L}}_i(\vq)
    :=
    U(\vb*{G}_{\mathrm{S}})\mathcal{S}_i(\vq).
    \label{eq_line_screw_definition}
\end{align}
The superscript \(\mathrm{L}\) denotes the restriction to a high-symmetry line.
This operator satisfies
\begin{align}
    \mathcal{S}^{\mathrm{L}}_i(\vq)
    H(\vq)
    \qty[\mathcal{S}^{\mathrm{L}}_i(\vq)]^{-1}
    =
    H(\vq)
    \label{eq_line_screw_constraint}
\end{align}
on the corresponding high-symmetry line.
The explicit expressions of \(\mathcal{S}^{\mathrm{L}}_i(\vq)\) are given in Table~\ref{table_line_screw}.

The eigenvalues of \(\mathcal{S}^{\mathrm{L}}_i(\vq)\) can be obtained from the algebra of the symmetry operations.
Since \((C_{2i})^2=1\), the relation \(C_{2i}\vq=\vq+\vb*{G}_{\mathrm{S}}\) gives
\begin{align}
    C_{2i}\vb*{G}_{\mathrm{S}}
    =
    -\vb*{G}_{\mathrm{S}}.
\end{align}
The screw operator and the reciprocal-lattice shift satisfy
\begin{align}
    \mathcal{S}_i(\vk)U(\vb*{G})
    =
    U(C_{2i}\vb*{G})\mathcal{S}_i(\vk+\vb*{G}),
\end{align}
which expresses the consistency between applying the reciprocal-lattice shift before and after the screw operation.
Using these relations, the square of \(\mathcal{S}^{\mathrm{L}}_i(\vq)\) reduces to the representation of the squared screw operation,
\begin{align}
    \qty[
        \mathcal{S}^{\mathrm{L}}_i(\vq)
    ]^2
    =
    \mathcal{S}_i(C_{2i}\vq)
    \mathcal{S}_i(\vq)
    =
    -e^{-\mathrm{i}\tilde{q}_i}
    I,
    \label{eq_line_screw_square}
\end{align}
where \(I\) denotes the \(8\times8\) identity matrix.
The minus sign comes from the spin part of the twofold rotation.
Equation~\eqref{eq_line_screw_square} shows that the eigenvalues of \(\mathcal{S}^{\mathrm{L}}_i(\vq)\) can be written as \(\chi\,\mathrm{i}\,e^{-\mathrm{i}\tilde{q}_i/2}\) with \(\chi=\pm1\).
The sign \(\chi\) labels the \(+\) and \(-\) screw sectors.
The common phase factor \(\mathrm{i}\,e^{-\mathrm{i}\tilde{q}_i/2}\) does not affect the distinction between these sectors.
The screw symmetry prevents hybridization between states in different screw sectors, as shown in Fig.~\ref{fig_band_waSOC}.

Kramers degeneracies on the high-symmetry lines are analyzed using antiunitary operators.
For each line where such an operator exists, we consider a screw operation whose rotational part maps \(-\vq\) to an equivalent wave vector,
\begin{align}
    C_{2j}(-\vq)
    =
    \vq+\vb*{G}_{\mathrm{A}} ,
\end{align}
where \(j=x,y,z\), and \(\vb*{G}_{\mathrm{A}}\) is the reciprocal lattice vector associated with the high-symmetry line and screw operation considered.
Combining this screw operation with time reversal gives an antiunitary operator acting within the space of Bloch states at \(\vq\),
\begin{align}
    \mathcal{A}(\vq)
    :=
    U(\vb*{G}_{\mathrm{A}})\mathcal{S}_j(-\vq)\Theta.
    \label{eq_line_antiunitary_definition}
\end{align}
For the operators listed in Table~\ref{table_line_screw}, one finds on the corresponding high-symmetry lines
\begin{align}
    \qty[\mathcal{A}(\vq)]^2
    =
    -I,
    \label{eq_line_antiunitary_constraint}
\end{align}
and
\begin{align}
    \mathcal{A}(\vq)H(\vq)[\mathcal{A}(\vq)]^{-1}
    =
    H(\vq).
    \label{eq_line_antiunitary_symmetry}
\end{align}
These relations enforce a Kramers degeneracy.
High-symmetry lines for which no such operator is present are marked by ``--'' in Table~\ref{table_line_screw}.

\begin{table}
    \centering
    \begin{tabular*}{\columnwidth}{@{\extracolsep{\fill}}cccc@{}}
        \hline\hline
        Symmetry line
        & \(\mathcal{S}^{\mathrm{L}}_i(\vq)\)
        & \(\mathcal{A}(\vq)\)
        & \(\xi\)
        \\
        \hline
        \(\Gamma\)-X
        & \(\mathcal{S}_x(\vq)\)
        & --
        & --
        \\
        X-S
        & \(U(-\vb*{a}^{\ast})\mathcal{S}_y(\vq)\)
        & \(U(-\vb*{a}^{\ast})\mathcal{S}_x(-\vq)\Theta\)
        & \(-1\)
        \\
        S-Y
        & \(U(-\vb*{b}^{\ast})\mathcal{S}_x(\vq)\)
        & \(U(-\vb*{b}^{\ast})\mathcal{S}_y(-\vq)\Theta\)
        & \(-1\)
        \\
        Y-\(\Gamma\)
        & \(\mathcal{S}_y(\vq)\)
        & --
        & --
        \\
        \(\Gamma\)-Z
        & \(\mathcal{S}_z(\vq)\)
        & --
        & --
        \\
        Z-U
        & \(U(-\vb*{c}^{\ast})\mathcal{S}_x(\vq)\)
        & \(U(-\vb*{c}^{\ast})\mathcal{S}_z(-\vq)\Theta\)
        & \(-1\)
        \\
        U-R
        & \(U(-\vb*{a}^{\ast}-\vb*{c}^{\ast})\mathcal{S}_y(\vq)\)
        & \(U(-\vb*{a}^{\ast})\mathcal{S}_x(-\vq)\Theta\)
        & \(+1\)
        \\
        R-T
        & \(U(-\vb*{b}^{\ast}-\vb*{c}^{\ast})\mathcal{S}_x(\vq)\)
        & \(U(-\vb*{b}^{\ast})\mathcal{S}_y(-\vq)\Theta\)
        & \(+1\)
        \\
        T-Z
        & \(U(-\vb*{c}^{\ast})\mathcal{S}_y(\vq)\)
        & \(U(-\vb*{c}^{\ast})\mathcal{S}_z(-\vq)\Theta\)
        & \(-1\)
        \\
        \hline\hline
    \end{tabular*}
    \caption{
    Screw operators \(\mathcal{S}^{\mathrm{L}}_i(\vq)\) and antiunitary operators \(\mathcal{A}(\vq)\) on the high-symmetry lines.
    The sign \(\xi\) characterizes the screw sector of the degenerate partner, where \(\xi=+1\) and \(\xi=-1\) denote the same and opposite screw sectors, respectively.
    }
    \label{table_line_screw}
\end{table}

\begin{table}
    \centering
    \begin{tabular*}{\columnwidth}{@{\extracolsep{\fill}}cccc@{}}
        \hline\hline
        \shortstack{Symm.\\point}
        & \(\mathcal{S}^{\mathrm{L}}_i(\vb*{Q})\)
        & \(\mathcal{A}_{+}(\vb*{Q})\)
        & \(\mathcal{A}_{-}(\vb*{Q})\)
        \\
        \hline
        S
        & \(U(-\vb*{a}^{\ast})\mathcal{S}_y(\vb*{Q})\)
        & \(U(-\vb*{b}^{\ast})\mathcal{S}_y(-\vb*{Q})\Theta\)
        & \(U(-\vb*{a}^{\ast})\mathcal{S}_x(-\vb*{Q})\Theta\)
        \\
        U
        & \(U(-\vb*{c}^{\ast})\mathcal{S}_x(\vb*{Q})\)
        & \(U(-\vb*{a}^{\ast})\mathcal{S}_x(-\vb*{Q})\Theta\)
        & \(U(-\vb*{c}^{\ast})\mathcal{S}_z(-\vb*{Q})\Theta\)
        \\
        T
        & \(U(-\vb*{c}^{\ast})\mathcal{S}_y(\vb*{Q})\)
        & \(U(-\vb*{b}^{\ast})\mathcal{S}_y(-\vb*{Q})\Theta\)
        & \(U(-\vb*{c}^{\ast})\mathcal{S}_z(-\vb*{Q})\Theta\)
        \\
        \hline\hline
    \end{tabular*}
    \caption{
    Screw operators \(\mathcal{S}^{\mathrm{L}}_i(\vb*{Q})\) and antiunitary operators \(\mathcal{A}_{\pm}(\vb*{Q})\) at the high-symmetry points \(\mathrm{S}\), \(\mathrm{U}\), and \(\mathrm{T}\).
    }
    \label{table_point_fourfold}
\end{table}

On a high-symmetry line with a twofold degeneracy, the degenerate partners can be classified according to whether they belong to the same or opposite screw sectors.
We denote an eigenstate of \(H(\vq)\) in the screw sector \(\chi=\pm1\) by \(\ket{\chi,\vq}\),
\begin{align}
    \mathcal{S}^{\mathrm{L}}_i(\vq)
    \ket{\chi,\vq}
    =
    \chi\,\mathrm{i}\,e^{-\mathrm{i}\tilde{q}_i/2}
    \ket{\chi,\vq}.
\end{align}
Since \(\mathcal{A}(\vq)\) satisfies Eqs.~\eqref{eq_line_antiunitary_constraint} and \eqref{eq_line_antiunitary_symmetry}, the state
\(\mathcal{A}(\vq)\ket{\chi,\vq}\) is a degenerate partner of
\(\ket{\chi,\vq}\).
To determine its screw sector, it is useful to introduce the projector
\begin{align}
    P_{\chi}(\vq)
    =
    \frac{1}{2}
    \qty(
    I
    -
    \chi\,\mathrm{i}e^{\mathrm{i}\tilde{q}_i/2}
    \mathcal{S}^{\mathrm{L}}_i(\vq)
    ).
\end{align}
Using the line-dependent matrix representations of \(\mathcal{S}^{\mathrm{L}}_i(\vq)\) and \(\mathcal{A}(\vq)\) listed in Table~\ref{table_line_screw}, we determine \(\xi\) from
\begin{align}
    \mathcal{A}(\vq)
    P_{\chi}(\vq)
    [\mathcal{A}(\vq)]^{-1}
    =
    P_{\xi\chi}(\vq).
\end{align}
The value of \(\xi\) for each high-symmetry line is given in Table~\ref{table_line_screw}.
For \(\xi=+1\) and \(\xi=-1\), \(\mathcal{A}(\vq)\) maps a state to its degenerate partner in the same and opposite screw sectors, respectively.

At the high-symmetry points \(\mathrm{S}\), \(\mathrm{U}\), and \(\mathrm{T}\), the fourfold degeneracy can be understood by applying the same screw-sector decomposition at the endpoint of the corresponding high-symmetry line.
For each point, we use the screw operator \(\mathcal{S}^{\mathrm{L}}_i(\vb*{q})\) on that line and evaluate it at \(\vb*{q}=\vb*{Q}\), where \(\vb*{Q}\) denotes the wave vector of the high-symmetry point.
The corresponding projector \(P_{\chi}(\vb*{Q})\) is obtained from \(P_{\chi}(\vq)\) by replacing \(\vq\) with \(\vb*{Q}\).
For the matrix representations listed in Table~\ref{table_point_fourfold}, the two antiunitary operators \(\mathcal{A}_{+}(\vb*{Q})\) and \(\mathcal{A}_{-}(\vb*{Q})\) satisfy
\begin{align}
    [\mathcal{A}_{\pm}(\vb*{Q})]^2
    =
    -I,
\end{align}
and
\begin{align}
    &\mathcal{A}_{\pm}(\vb*{Q})
    H(\vb*{Q})
    [\mathcal{A}_{\pm}(\vb*{Q})]^{-1}
    =
    H(\vb*{Q}), \\
    &\mathcal{A}_{\pm}(\vb*{Q})
    P_{\chi}(\vb*{Q})
    [\mathcal{A}_{\pm}(\vb*{Q})]^{-1}
    =
    P_{\pm\chi}(\vb*{Q}).
\end{align}
Therefore, for an eigenstate \(\ket{\chi,\vb*{Q}}\), the four states
\(\ket{\chi,\vb*{Q}}\),
\(\mathcal{A}_{+}(\vb*{Q})\ket{\chi,\vb*{Q}}\),
\(\mathcal{A}_{-}(\vb*{Q})\ket{\chi,\vb*{Q}}\), and
\(\mathcal{A}_{+}(\vb*{Q})\mathcal{A}_{-}(\vb*{Q})\ket{\chi,\vb*{Q}}\)
are degenerate.
The first two states belong to the screw sector \(\chi\), while the last two belong to the screw sector \(-\chi\).
Since the two screw sectors are orthogonal, these four states are linearly independent, proving the fourfold degeneracy at
\(\mathrm{S}\), \(\mathrm{U}\), and \(\mathrm{T}\).

The same symmetry analysis also provides a simple criterion for the spin components that can remain finite after averaging over a degenerate subspace.
The spin operator \(O_j^s=\rho_0\sigma_0s_j/2\), with \(j=1,2,3\), transforms under \(\mathcal{S}^{\mathrm{L}}_i(\vq)\) as
\begin{align}
    \mathcal{S}^{\mathrm{L}}_i(\vq)
    O^s_j
    \qty[
        \mathcal{S}^{\mathrm{L}}_i(\vq)
    ]^{-1}
    =
    \begin{cases}
        O^s_j, & j=i, \\
        -O^s_j, & j\neq i.
    \end{cases}
\end{align}
The site-space part of \(\mathcal{S}^{\mathrm{L}}_i(\vq)\) commutes with \(O^s_j\), whereas the spin part determines whether \(O^s_j\) is even or odd under \(\mathcal{S}^{\mathrm{L}}_i(\vq)\), as follows from \(\{s_i,s_j\}=2\delta_{ij}s_0\).
Within a degenerate subspace invariant under \(\mathcal{S}^{\mathrm{L}}_i(\vq)\), the averaged expectation value of an operator that is odd under \(\mathcal{S}^{\mathrm{L}}_i(\vq)\) must vanish.
Therefore, the averaged expectation value of \(O^s_j\) can remain finite only for \(j=i\), while it vanishes for \(j\neq i\).
At the time-reversal-invariant high-symmetry points, the degenerate subspace is also invariant under \(\Theta\).
Since all spin components are odd under time reversal, \(\Theta O^s_j\Theta^{-1}=-O^s_j\), the averaged expectation value of \(O^s_j\) vanishes for all \(j\).
This criterion accounts for the line-dependent spin components and their vanishing at the high-symmetry points, as shown in Fig.~\ref{fig_spin}.

\section{Conclusion}
\label{sec_conclusion}

We have constructed a symmetry-adapted effective tight-binding model for a chiral one-dimensional organic-inorganic lead halide perovskite with space group \(P2_12_12_1\) (No. 19).
The model is built from a single effective orbital on each of the four relevant sites in the primitive unit cell.
The resulting four-dimensional site space is represented by layer and in-plane sublattice degrees of freedom.
Together with spin, these degrees of freedom define an \(8\times8\) effective Hamiltonian that respects the \(P2_12_12_1\) space-group symmetry and time-reversal symmetry.

By fitting the hopping parameters separately for the conduction and valence bands, the effective Hamiltonian reproduces the overall band dispersions obtained from DFT calculations.
The spin-dependent hopping terms reproduce the band-edge spin splittings and capture the leading spin-polarization patterns found in the DFT results.
These results show that the band-edge electronic structure can be captured by a restricted set of symmetry-adapted hopping terms.

We have also analyzed the degeneracies and band crossings in terms of the screw symmetries of \(P2_12_12_1\).
By adding further symmetry-adapted spin-dependent terms, we separated accidental degeneracies of the effective Hamiltonian from degeneracies enforced by screw symmetries and time-reversal symmetry.
The screw-eigenvalue classification on high-symmetry lines explains which crossing branches can hybridize and form anticrossings, and which crossings remain protected by screw symmetry.
This analysis gives an explicit matrix representation of the screw-sector classification and the symmetry-enforced degeneracies.

The present model provides a low-energy microscopic description for understanding how chiral crystal structure and symmetry-adapted spin-dependent terms determine band-edge electronic properties in chiral lead halide perovskites.
By resolving the distinct roles of the symmetry-adapted spin-dependent terms, including the different contributions of the \(\lambda^{\pm}_{x}\) and \(\lambda^{\pm}_{y}\) terms to the band-edge spin splitting and spin-polarization patterns, the model provides microscopic insight into the origin of these features that cannot be directly identified from the DFT band structure alone.
The compact and symmetry-transparent form of the effective Hamiltonian provides a framework for systematically analyzing how microscopic hopping processes and spin-dependent terms contribute to chirality-dependent responses.
This effective-model approach complements DFT and Wannier-based methods by providing a tractable starting point for developing theories of spin-dependent, optical, and transport responses in chiral lead halide perovskites.

\section*{Acknowledgments}

This work was supported by JSPS KAKENHI (Grants No.~JP23K17659, No.~JP25K00963, No.~JP26K06985, and No.~JP26H00378), JST CREST (Grant No.~JPMJCR23A1), and Waseda University Grant for Special Research Projects (No.~2026C-668).

\appendix
\input{appendix}

\bibliography{ref}

\end{document}

%% file: appendix.tex
\appendix

\section{Screw operations in the Bloch basis}
\label{app_screw_representation}

\begin{table}[h!]
    \centering
    {
    \renewcommand{\arraystretch}{1.4}
    \begin{tabular*}{\columnwidth}{@{\extracolsep{\fill}}ccc@{}}
        \hline\hline
        Screw 
        & Site permutation
        & Site-space matrix
        \\
        \hline
        \(S_x\)
        & \(A\leftrightarrow B,\ C\leftrightarrow D\)
        & \(\rho_0\sigma_1\)
        \\
        \(S_y\)
        & \(A\leftrightarrow C,\ B\leftrightarrow D\)
        & \(\rho_1\sigma_0\)
        \\
        \(S_z\)
        & \(A\leftrightarrow D,\ B\leftrightarrow C\)
        & \(\rho_1\sigma_1\)
        \\
        \hline\hline
    \end{tabular*}
    }
    \caption{
    Screw operations, site permutations, and matrix representations in the site space.
    }
    \label{table_app_screw_site}
\end{table}

This appendix derives the matrix representations of the screw operations in the Bloch basis introduced in Sec.~\ref{sec_model}.
For the screw operation \(S_i\), we denote its real-space translation vector by \(\vb*{t}_i\).
Under this operation, the position of site \(X\) is mapped to a site equivalent, up to a Bravais lattice vector, to one of the four sites in the unit cell,
\begin{align}
    S_i:\vb*{\tau}_{X}\mapsto C_{2i}\vb*{\tau}_{X}+\vb*{t}_i
    =
    \vb*{L}_{iX}+\vb*{\tau}_{X^\prime},
    \label{eq_app_site_decomposition}
\end{align}
where \(X^\prime\) is the site to which \(X\) is mapped by \(S_i\), and \(\vb*{L}_{iX}\) is a Bravais lattice vector.
The action on a localized basis state is
\begin{align}
    S_i\ket{\vb*{R}+\vb*{\tau}_{X},s}
    =
    \sum_{s^\prime}
    \qty(-\mathrm{i}s_i)_{s^\prime s}
    \ket{
        C_{2i}\vb*{R}
        +
        \vb*{L}_{iX}
        +
        \vb*{\tau}_{X^\prime},
        s^\prime
    },
    \label{eq_app_localized_screw_action}
\end{align}
where \(-\mathrm{i}s_i\) is the spin part of a twofold rotation about the \(i\) axis.
Substituting Eq.~\eqref{eq_app_localized_screw_action} into the Bloch basis gives
\begin{align}
    S_i\ket{\vk,X,s}
    =
    e^{-\mathrm{i}(C_{2i}\vk)\cdot\vb*{t}_i}
    \sum_{s^\prime}
    \qty(-\mathrm{i}s_i)_{s^\prime s}
    \ket{C_{2i}\vk,X^\prime,s^\prime}.
    \label{eq_app_screw_action_general}
\end{align}
In deriving this expression, the shift by the Bravais vector \(\vb*{L}_{iX}\) is absorbed by relabeling the Bravais-lattice summation variable.
The site permutations and the corresponding matrix representations in the site space are summarized in Table~\ref{table_app_screw_site}.
The phase factors are given by
\begin{align}
        e^{-\mathrm{i}(C_{2x}\vk)\cdot\vb*{t}_x}
        &=
        e^{-\mathrm{i}(\tilde{k}_x-\tilde{k}_y)/2}, \\
        e^{-\mathrm{i}(C_{2y}\vk)\cdot\vb*{t}_y}
        &=
        e^{-\mathrm{i}(\tilde{k}_y-\tilde{k}_z)/2}, \\
        e^{-\mathrm{i}(C_{2z}\vk)\cdot\vb*{t}_z}
        &=
        e^{-\mathrm{i}(\tilde{k}_z-\tilde{k}_x)/2}.
\end{align}
Using the above equations and Table~\ref{table_app_screw_site}, we obtain the screw-operation matrices in Eqs.~\eqref{eq_screw_x_full}--\eqref{eq_screw_z_full}.

%% file: ref.bib
@article{Perdew1996,
  title = {Generalized Gradient Approximation Made Simple},
  author = {Perdew, John P. and Burke, Kieron and Ernzerhof, Matthias},
  journal = {Phys. Rev. Lett.},
  volume = {77},
  issue = {18},
  pages = {3865--3868},
  numpages = {0},
  year = {1996},
  month = {Oct},
  publisher = {American Physical Society},
  doi = {10.1103/PhysRevLett.77.3865}
}

@ARTICLE{Giannozzi2009,
  title     = "{QUANTUM ESPRESSO}: A modular and open-source software project for quantum simulations of materials",
  author    = "Giannozzi, Paolo and Baroni, Stefano and Bonini, Nicola and Calandra, Matteo and Car, Roberto and Cavazzoni, Carlo and Ceresoli, Davide and Chiarotti, Guido L and Cococcioni, Matteo and Dabo, Ismaila and Dal Corso, Andrea and de Gironcoli, Stefano and Fabris, Stefano and Fratesi, Guido and Gebauer, Ralph and Gerstmann, Uwe and Gougoussis, Christos and Kokalj, Anton and Lazzeri, Michele and Martin-Samos, Layla and Marzari, Nicola and Mauri, Francesco and Mazzarello, Riccardo and Paolini, Stefano and Pasquarello, Alfredo and Paulatto, Lorenzo and Sbraccia, Carlo and Scandolo, Sandro and Sclauzero, Gabriele and Seitsonen, Ari P and Smogunov, Alexander and Umari, Paolo and Wentzcovitch, Renata M",
  journal   = "J. Phys.: Condens. Matter",
  publisher = "IOP Publishing",
  volume    = 21,
  number    = 39,
  pages     = 395502,
  month     = sep,
  year      = 2009,
  doi       = "10.1088/0953-8984/21/39/395502"
}

@ARTICLE{Giannozzi2017,
  title     = {Advanced capabilities for materials modelling with {Quantum ESPRESSO}},
  author    = {Giannozzi, Paolo and Andreussi, Oliviero and Brumme, Thomas and
               Bunau, Olivier and Buongiorno Nardelli, Marco and Calandra, Matteo and
               Car, Roberto and Cavazzoni, Carlo and Ceresoli, Davide and
               Cococcioni, Matteo and Colonna, Nicola and Carnimeo, Ivan and
               Dal Corso, Andrea and de Gironcoli, Stefano and Delugas, Pietro and
               DiStasio, Jr., Robert A. and Ferretti, Andrea and Floris, Andrea and
               Fratesi, Guido and Fugallo, Giorgia and Gebauer, Ralph and
               Gerstmann, Uwe and Giustino, Feliciano and Gorni, Tommaso and
               Jia, Jiayu and Kawamura, Mitsuaki and Ko, Hyungjun-Yu and
               Kokalj, Anton and K{\"u}{\c c}{\"u}kbenli, Emine and Lazzeri, Michele and
               Marsili, Matteo and Marzari, Nicola and Mauri, Francesco and
               Nguyen, Nguyen Lan and Nguyen, Huy-Viet and
               Otero-de-la-Roza, Alberto and Paulatto, Lorenzo and
               Ponc{\'e}, Samuel and Rocca, Dario and Sabatini, Riccardo and
               Santra, Biswajit and Schlipf, Martin and Seitsonen, Ari P. and
               Smogunov, Alexander and Timrov, Iurii and Thonhauser, Timo and
               Umari, Paolo and Vast, Nathalie and Wu, Xifan and Baroni, Stefano},
  journal   = {J. Phys.: Condens. Matter},
  publisher = {IOP Publishing},
  volume    = {29},
  number    = {46},
  pages     = {465901},
  month     = nov,
  year      = {2017},
  doi       = {10.1088/1361-648X/aa8f79}
}

@ARTICLE{Herring1937,
  title   = "Accidental Degeneracy in the Energy Bands of Crystals",
  author  = "Herring, Conyers",
  journal = "Phys. Rev.",
  volume  = "52",
  pages   = "365--373",
  year    = "1937",
  doi     = "10.1103/PhysRev.52.365"
}

@ARTICLE{Konig1997,
  title   = "Electronic level degeneracy in nonsymmorphic periodic or aperiodic crystals",
  author  = "K{\"o}nig, A. and Mermin, N. D.",
  journal = "Phys. Rev. B",
  volume  = "56",
  pages   = "13607--13610",
  year    = "1997",
  doi     = "10.1103/PhysRevB.56.13607"
}

@ARTICLE{Konig1999,
  title   = "Screw rotations and glide mirrors: Crystallography in Fourier space",
  author  = "K{\"o}nig, Anja and Mermin, N. David",
  journal = "Proc. Natl. Acad. Sci.",
  volume  = "96",
  number  = "7",
  pages   = "3502--3506",
  year    = "1999",
  doi     = "10.1073/pnas.96.7.3502"
}

@BOOK{Bradley2010,
  title={The mathematical theory of symmetry in solids: representation theory for point groups and space groups},
  author={Bradley, Christopher and Cracknell, Arthur},
  year={2010},
  publisher={Oxford University Press, New York}
}

@ARTICLE{Furusaki2017,
  title     = "Weyl points and Dirac lines protected by multiple screw rotations",
  author    = "Furusaki, Akira",
  journal   = "Science Bulletin",
  publisher = "Elsevier BV",
  volume    = 62,
  number    = 11,
  pages     = "788--794",
  month     = jun,
  year      = 2017,
  doi       = "10.1016/j.scib.2017.05.014"
}

@article{Chang2018,
  title = {Topological quantum properties of chiral crystals},
  author = {Chang, Guoqing and Wieder, Benjamin J. and Schindler, Frank and Sanchez, Daniel S. and Belopolski, Ilya and Huang, Shin-Ming and Singh, Bahadur and Wu, Di and Chang, Tay-Rong and Neupert, Titus and Xu, Su-Yang and Lin, Hsin and Hasan, M. Zahid},
  journal = {Nature Mater},
  volume = {17},
  pages = {978--985},
  year = {2018},
  doi = {10.1038/s41563-018-0169-3}
}

@ARTICLE{Leonhardt2021,
  title     = "Symmetry-enforced topological band crossings in orthorhombic crystals: Classification and materials discovery",
  author    = "Leonhardt, Andreas and Hirschmann, Moritz M and Heinsdorf, Niclas and Wu, Xianxin and Fabini, Douglas H and Schnyder, Andreas P",
  journal   = "Phys. Rev. Materials",
  publisher = "American Physical Society (APS)",
  volume    = 5,
  number    = 12,
  pages     = 124202,
  month     = dec,
  year      = 2021,
  doi       = "10.1103/PhysRevMaterials.5.124202"
}

@BOOK{Vanderbilt2018,
  title     = "Berry Phases in Electronic Structure Theory: Electric Polarization, Orbital Magnetization and Topological Insulators",
  author    = "Vanderbilt, David",
  publisher = "Cambridge University Press, Cambridge, England",
  year      = 2018
}

@ARTICLE{EsteveParedes2023,
  title   = "A comprehensive study of the velocity, momentum and position matrix elements for Bloch states: application to a local orbital basis",
  author  = "Esteve-Paredes, Juan Jos{\'e} and Palacios, Juan Jos{\'e}",
  journal = "SciPost Phys. Core",
  volume  = 6,
  pages   = "002",
  year    = 2023,
  doi     = "10.21468/SciPostPhysCore.6.1.002"
}

@MISC{WannierToolsAtomicGauge,
  title        = "WannierTools: Tight-binding model",
  howpublished = "\url{https://www.wanniertools.org/theory/tight-binding-model/}",
  note         = "Accessed July 2026"
}

@ARTICLE{Long2020,
  title     = "Chiral-perovskite optoelectronics",
  author    = "Long, Guankui and Sabatini, Randy and Saidaminov, Makhsud I. and Lakhwani, Girish and Rasmita, Abdullah and Liu, Xiaogang and Sargent, Edward H. and Gao, Weibo",
  journal   = "Nat. Rev. Mater.",
  publisher = "Springer Science and Business Media LLC",
  volume    = 5,
  number    = 6,
  pages     = "423--439",
  month     = mar,
  year      = 2020,
  doi       = "10.1038/s41578-020-0181-5"
}

@ARTICLE{Pietropaolo2022,
  title     = "Rationalizing the design and implementation of chiral hybrid perovskites",
  author    = "Pietropaolo, Adriana and Mattoni, Alessandro and Pica, Giovanni and Fortino, Mariagrazia and Schifino, Gioacchino and Grancini, Giulia",
  journal   = "Chem",
  publisher = "Elsevier BV",
  volume    = 8,
  number    = 5,
  pages     = "1231--1253",
  month     = may,
  year      = 2022,
  doi       = "10.1016/j.chempr.2022.01.014"
}

@ARTICLE{Ishii2025,
  title     = "Giant bulk photovoltaic effect in a chiral polar crystal based on
               helical one-dimensional lead halide perovskites",
  author    = "Ishii, Ayumi and Sone, Ryohei and Yamada, Tomohide and Noto,
               Mizuki and Suzuki, Hikari and Nakamura, Daiki and Murata, Kei and
               Shiga, Takuya and Ishii, Kazuyuki and Nihei, Masayuki",
  journal   = "Angew. Chem. Int. Ed.",
  publisher = "Wiley",
  volume    = 64,
  number    = 16,
  pages     = "e202424391",
  year      = 2025,
  doi       = "10.1002/anie.202424391"
}

@ARTICLE{Billing2003,
  title   = "Bis[({S})-$\beta$-phenethylammonium] tribromoplumbate({II})",
  author  = "Billing, David G. and Lemmerer, Andreas",
  journal = "Acta Crystallogr. Sect. E Struct. Rep. Online",
  volume  = 59,
  number  = 6,
  pages   = "m381--m383",
  year    = 2003,
  doi     = "10.1107/S1600536803010985"
}

@ARTICLE{Chen2019,
  title     = "Circularly polarized light detection using chiral hybrid perovskite",
  author    = "Chen, Chao and Gao, Liang and Gao, Wanru and Ge, Cong and Du, Xinyuan and Li, Zha and Yang, Ying and Niu, Guangda and Tang, Jiang",
  journal   = "Nat. Commun.",
  volume    = 10,
  number    = 1,
  pages     = "1927",
  year      = 2019,
  doi       = "10.1038/s41467-019-09942-z"
}

@ARTICLE{Ishii2020,
  title     = "Direct detection of circular polarized light in helical {1D} perovskite-based photodiode",
  author    = "Ishii, Ayumi and Miyasaka, Tsutomu",
  journal   = "Sci. Adv.",
  volume    = 6,
  number    = 46,
  pages     = "eabd3274",
  year      = 2020,
  doi       = "10.1126/sciadv.abd3274"
}

@ARTICLE{Hu2020,
  title     = "A chiral switchable photovoltaic ferroelectric {1D} perovskite",
  author    = "Hu, Yang and Florio, Fred and Chen, Zhizhong and Phelan, W. Adam and Siegler, Maxime A. and Zhou, Zhe and Guo, Yuwei and Hawks, Ryan and Jiang, Jie and Feng, Jing and Zhang, Lifu and Wang, Baiwei and Wang, Yiping and Gall, Daniel and Palermo, Edmund F. and Lu, Zonghuan and Sun, Xin and Lu, Toh-Ming and Zhou, Hua and Ren, Yang and Wertz, Esther and Sundararaman, Ravishankar and Shi, Jian",
  journal   = "Sci. Adv.",
  volume    = 6,
  number    = 9,
  pages     = "eaay4213",
  year      = 2020,
  doi       = "10.1126/sciadv.aay4213"
}

@ARTICLE{Makhija2025,
  title     = {Piezoelectricity in a mixture of chiral {1D} hybrid lead bromide
               and iodide systems},
  author    = {Makhija, Urmila and Kushwaha, Vikash and Prajesh, Neetu and
               Nag, Angshuman and Boomishankar, Ramamoorthy},
  journal   = {J. Mater. Chem. C},
  publisher = {Royal Society of Chemistry},
  volume    = {13},
  number    = {46},
  pages     = {23037--23043},
  year      = {2025},
  doi       = {10.1039/D5TC02798E}
}

@ARTICLE{Wei2021,
  title     = "Giant spin splitting in chiral perovskites based on local electrical field engineering",
  author    = "Wei, Qi and Zhang, Qingyun and Xiang, Longjun and Zhang, Shihao and Liu, Jianpeng and Yang, Xiaoyu and Ke, Youqi and Ning, Zhijun",
  journal   = "J. Phys. Chem. Lett.",
  publisher = "American Chemical Society (ACS)",
  volume    = 12,
  number    = 28,
  pages     = "6492--6498",
  month     = jul,
  year      = 2021,
  doi       = "10.1021/acs.jpclett.1c01675"
}

@ARTICLE{Xiao2024,
  title     = "Strain-amplified exciton chirality in organic-inorganic hybrid materials",
  author    = "Xiao, Jin and Zheng, Haofeng and Liu, Yanan and Fang, Li and Li, Jing and Kim, Jongchan and Wang, Yanlong and Liu, Qi and Ma, Xuyu and Hou, Shaocong",
  journal   = "Phys. Rev. Lett.",
  publisher = "American Physical Society (APS)",
  volume    = 133,
  number    = 5,
  pages     = 056903,
  month     = aug,
  year      = 2024,
  doi       = "10.1103/PhysRevLett.133.056903"
}

@ARTICLE{Furukawa2026,
  title   = {First-principles calculations of spin-split bands in chiral
             hybrid organic-inorganic perovskites
             {($R/S$-PEA)PbI$_3$} and {($R/S$-NEA)PbI$_3$}},
  author  = {Furukawa, Tetsuya and Nakano, Kazushi and Suzuki, Youta and
             Kaneko, Takumi and Ishii, Ayumi and Itou, Tetsuaki},
  journal = {arXiv:2607.00933},
  year    = {2026},
  doi     = {10.48550/arXiv.2607.00933}
}
